\documentclass{article}

\usepackage{arxiv}

\usepackage[utf8]{inputenc} 
\usepackage[T1]{fontenc}    
\usepackage{hyperref}       
\usepackage{url}            
\usepackage{booktabs}       
\usepackage{amsfonts}       
\usepackage{nicefrac}       
\usepackage{microtype}      
\usepackage{cleveref}       
\usepackage{lipsum}         
\usepackage{graphicx}
\usepackage{natbib}
\usepackage{doi}

\usepackage{algorithm}
\usepackage{algorithmic}

\usepackage{subcaption}
\title{A K-means Inspired Solution Framework for Large-Scale Multi-Traveling Salesman Problems}

\newif\ifuniqueAffiliation
\uniqueAffiliationtrue

\author{%
  \href{https://orcid.org/0009-0001-4168-4758}{%
    \includegraphics[scale=0.06]{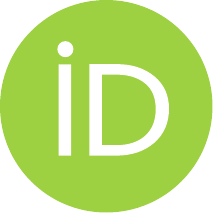}\hspace{1mm}Xiubin Chen%
  }\\
  School of Astronautics\\
  Harbin Institute of Technology\\
  Harbin, China \\
  \texttt{24S004020@stu.hit.edu.cn} \\
}

\renewcommand{\headeright}{Technical Report}
\renewcommand{\undertitle}{Technical Report}
\renewcommand{\shorttitle}{\textit{arXiv} A K-means Inspired Solution Framework}

\hypersetup{
pdftitle={A K-means Inspired Solution Framework for Large-Scale Multi-Traveling Salesman Problems},
pdfsubject={Computer Science – Robotics},
pdfauthor={Xiubin Chen},
pdfkeywords={Multi-agent systems, MTSP, Task allocation, K-means clustering, Large-scale coordination},
}

\begin{document}
\maketitle

\begin{abstract}
    The Multi-Traveling Salesman Problem (MTSP) is a commonly used mathematical model for
    multi-agent task allocation. However, as the number of agents and task targets increases,
    existing optimization-based methods often incur prohibitive computational costs, posing significant
    challenges to large-scale coordination in unmanned systems. To address this issue, this paper
    proposes a K-means-inspired task allocation framework that reformulates the MTSP as a spatially constrained
    classification process. By leveraging spatial coherence, the proposed method enables fast estimation of path
    costs and efficient task grouping, thereby fundamentally reducing overall computational complexity.
    Extensive simulation results demonstrate that the framework can maintain high solution quality
    even in extremely large-scale scenarios-for instance, in tasks involving 1000 agents and 5000 targets.
    The findings indicate that this "cluster-then-route" decomposition strategy offers an efficient
    and reliable solution for large-scale multi-agent task allocation.
\end{abstract}


\section{Introduction}

In the research and application of large-scale multi-agent unmanned systems, achieving efficient and reliable initial
task allocation is a fundamental problem in domains such as UAV swarms, multi-robot collaboration, and autonomous
vehicle fleets.\cite{ma2022intelligent} The Multi-Traveling Salesman Problem (MTSP) has become a widely used mathematical model for these
tasks, as it describes the cooperative visitation of multiple agents to multiple spatial targets, with each agent
returning to its starting point upon task completion.\cite{singh2016review} However, conaventional MTSP solution methods-typically based on
integer programming, heuristic search, or evolutionary algorithms-exhibit computational complexity that grows
exponentially with the number of agents and task locations. In large-scale unmanned swarm scenarios, where the number
of tasks reaches several thousands and the number of agents scales to hundreds or even thousand, existing approaches
often fail to produce high-quality solutions within acceptable time limits. This frequently forces designers to adopt
hierarchical or heavily rule-based strategies, which in turn limits the practicality and scalability of multi-agent
systems.

In ultra-large-scale settings, the main factors affecting the efficiency of MTSP solving include the vast task space,
competition among agents, and the highly coupled nature of global route costs. Although exising research attempts to
mitigate these challenges using divide-and conquer approaches, task partitioning, or graph simplification techniques,
it remains difficult to simultaneously achieve high computational efficiency and good solution quality. This challenge
is particularly evident in scenarios where task locations exhibit clear spatial distributions, yet the spatial
structure and regional characteristics are not sufficiently exploited. Thus, developing a task allocation framework
capable of rapidly capturing spatial task features, reducing the degree of coupling, and maintaining solution quality
under large-scale conditions is of significant importance.

To address these challenges, this paper proposes a K-means-inspired task allocation framework that formulates the MTSP
as a joint optimization problem of spatially coherent classification and local routing. Specifically, the algorithm
first performs a geometric decomposition of the task space based on the spatial relationship between tasks and agents,
thereby dividing the global problem into several spatially compact subproblems. Subsequently, utilizing a precomputed
global distance matrix, the framework efficiently solves a local Traveling Salesman Problem (TSP) within each
subregion, avoiding repeated distance computations and significantly improving routing efficiency. Building upon this
decomposition, a lexicographic neighborhood search mechanism if introduced, in which each task is allowed to migrate
only among a small set of nearly candidate clusters, effectively reducing the search space and further enhancing
optimization performance.

The proposed framework not only offers strong scalability but also exhibits good adaptability to scenario-specific
constraints such as obstacle avoidance and kinematic limits. Extensive simulation results demonstrate that the method
maintains stable performance even in ultra-large-scale scenarios involving 1000 agents and 5000 task locations,
indicating promising potential for practical deployment in real-world multi-agent systems.

\section{Problem Statement}

\subsection{Multi-Traveling Salesman Problem}

The Multi-Traveling Salesman Problem (MTSP) is a multi-agent extension of the classical Traveling Salesman Problem
(TSP), and is widely used to describe the task allocation process in which multiple agents cooperatively visit
spatial targets. The basic setting is as follows: givin a system with $k$ salesmen and $n$ spatially discrete task
locations, each task must be visited by exactly one salesman, and each salesman departs from a predetermined depot
and eventually returns to that depot. This modeling framework naturally fits typical unmanned platform scenarios in
which vehicles depart from a base, complete assigned tasks, and then return to the same base.

Let the set of agents be
\begin{equation}
    \mathcal{D} = \{ d_1, d_2, \dots, d_k \}, \qquad d_g \in \mathbb{R}^2,
\end{equation}
and the set of task locations be
\begin{equation}
    \mathcal{C} = \{ c_1, c_2, \dots, c_n \}, \qquad c_i \in \mathbb{R}^2.
\end{equation}

Each task must be assigned to one salesman. We define the assignment variable
\begin{equation}
    A_i \in \{1, 2, \dots, k\}, \qquad i = 1, \dots, n,
\end{equation}
which indicates that task $c_i$ is served by the $g$-th salesman with $g = A_i$. The corresponding task set of salesman
$g$ is then
\begin{equation}
    \mathcal{C}_g = \{ c_i \mid A_i = g \}.
\end{equation}

The complete route of salesman $g$ can be represented as a closed tour that starts from its depot, visits all assigned
tasks, and then returns to the depot:
\begin{equation}
    R_g = (d_g, \, c_{i_1}, \, c_{i_2}, \dots, c_{i_{m_g}}, \, d_g).
\end{equation}

The route cost is defined as the sum of distances between consecutive nodes along the tour. Depending on the
characteristics of the environment, different distance metrics can be used: in free space, Euclidean distance is
typically adopted, while in environments with obstacles, the shortest feasible path distance obtained by A$^{*}$
search is preferred. Accordingly, the route cost of salesman $g$ can be written in a unified form as
\begin{equation}
    L_g = \sum_{j=0}^{m_g} \mathrm{dist}\big(R_g[j],\, R_g[j+1]\big),
\end{equation}
where $\mathrm{dist}(\cdot, \cdot)$ denotes either the Euclidean distance or the A$^{*}$ distance.

\subsection{K-means Clustering Algorithm}

The K-means clustering algorithm is a classical prototype-based unsupervised learning method.\cite{ahmed2020k} Its core idea is to
identify $k$ cluster centers in a given dataset that can represent the underlying spatial structre, and to obtain a
compact and well-seperated clustering by minimizing the squared distance between each sample and the center of the
cluster to which it is assigned.

K-means obtains the cluster partition by optimizing the following objective function:
\begin{equation}
    \min_{\{S_g, \mu_g\}} \sum_{g=1}^k \sum_{x \in S_g} \| x- \mu_g \|_2^2,
\end{equation}
where $S_g$ denotes the set of samples assigned to the $g$-th cluster, and $\mu_g$ is the corresponding cluster
centroud. This objective function measures the compactness of clusters; its optimal solution corresponds to a spatial
partition that is geometrically tight and minimizes the mean squared error in a statistical sense.

To solve the above optimization problem, K-means adopts a typical alternating minimization scheme, in which the
following two steps are iterated until convergence:

\paragraph{Assignment step.}
Given the current cluster centroids $\{\mu_1, \dots, \mu_k\}$, each data point is assigned to the cluster whose centroid
is closest to it in terms of Euclidean distance:
\begin{equation}
    x \in S_g \quad \mathrm{iff} \quad g = \arg\min_h \| x - \mu_h \|_2.
\end{equation}
This step minimizes, inder fixed centroids, the squared distance from each sample to its assigned cluster center.
Geometrically, the cluster boundaries are determined by the perpendicular bisectors between centroids, leading to a
natural Voronoi-type partition of the space.

\paragraph{Update step.}
Given a fixed cluster assignment, the centroid of each cluster is recomputed as the arithmetic mean of all data points
within that cluster:
\begin{equation}
    \mu_g = \frac{1}{|S_g|} \sum_{x \in S_g} x.
\end{equation}
This update rule can be viewed as the closed form optimal solution of the K-means objective under a fixed partition.
Specifically, by solving
\begin{equation}
    \min_{\mu_g} \sum_{x \in S_g} \| x - \mu_g \|_2^2,
\end{equation}
one directly obtains that the optimal centroid must be the arithmetic mean of the points in $S_g$. As a result, each
update step ensures that the K-means objective function is monotonically non-increasing, thereby driving the algorithm
toward a stable local optimum.

\subsection{Computational Bottlenecks in Traditional Task Allocation}

In traditional solution methods for task allocation problems, the overall procedure is typically organized into four
stages: \emph{ordering-evaluation-selection-iteration}. Among these stages, the evaluation step is often the most
computationally demanding. Due to the presence of various practical constraints, dynamic factors,  and path feasibility
conditions, the evaluation module must compute route costs, check feasibility, and sometimes perform local optimization
for each candidate solution. The resulting computational complexity is highly dependent on the specific application
scenario, which makes it difficult to design a unified and generally efficient evaluation framework.

At the same time, the ordering step in search-based algorithms generates a large number if candidate task execution
sequences, and each sequence requires at least one call to the evaluation module for cost computation. Since the number
of possible orderings grows exponentially with the problem size, the evaluation module may be invoked tens of thousands
or even mullions of times, leading to a highly repetitive and "mechanical" computational burden.

However, for a task set composed jof agents and target points, once the combination of tasks assigned to each agent is
fixed, the optimal visiting order within that combination is in many cases determined or nearly unique, especially in
TSP-like scenarios with a clear tour structure. Therefore, instead of searching directly in the enormous space of
permutations, it is more reasonable to first determine the \emph{combination structure} of tasks and agents, and then
solve a local routing within a restricted subspace.

Following this idea, the traditional pipeline \emph{ordering-evaluation-selection-iteration} can be reformulated as
\emph{combination-local ordering and evaluation-selection-iteration}.

The direct benefit of this reformulation is a significant reduction in the number of outer-loop iterations, which
fundamentally decreases the number of calls to the evaluation module and thus effectively controls the overall
computational cost.

More concretely, when a task point $c_i$ is considered for assignment to agent $g$, the new route cost becomes
\begin{equation}
    L_g(\mathcal{C}_g \cup \{c_i\}).
\end{equation}

Thus, the assignment decision for task $c_i$ can be formalized as
\begin{equation}
    A_i^\star=\arg\min_{g \in \{1,\dots,k\}}    L_g(\mathcal{C}_g \cup \{c_i\}),
\end{equation}
i.e., each task point is assigned to the agent that yields the smallest resulting route cost after its inclusion.

On a global level, the overall combination structure of all task points can be expressed as the following optimization
problem:
\begin{equation}
    \min_{\{\mathcal{C}_g\}}
    \sum_{g=1}^k L_g(\mathcal{C}_g),
    \]
    where each task set $\mathcal{C}_g$ is uniquely determined by the decision variables
    \[
    A_i \in \{1,\dots,k\}, \qquad i = 1,\dots,n.
\end{equation}

\section{Proposed Task Allocation Framework}

In this section, we present the proposed task allocation framework in detail. The algorithm begins by constructing a
global distance matrix over the set of depots and task points, which serves as the basis for all subsequent route
cost evaluations. For any pair of points
\begin{equation}
    p_i, \, p_j \in \mathcal{D} \cup \mathcal{C},
\end{equation}
under the Euclidean metric, the distance matrix is defined as
\begin{equation}
    D_{ij} = \|p_i - p_j\|_2.
\end{equation}
Since route evaluation essentially depends on pairwise distances between nodes, precomputing the matrix $D$ allows all
later geometric computations to be reduced to constant-time table lookups, thereby significantly reducing the
computational cost of local TSP solving.

In the initial solution construction stage, the framework adopts a geometry-based heuristic a geometry-based heuristic
assignment strategy. First, each agent is guaranteed to receive at least one task point. Then, the remaining task points
are assigned to depots according to a nearest-neighbor rule:
\begin{equation}
    A_i = \arg\min_{g} \|c_i - d_g\|_2,
\end{equation}
where $A_i$ denotes the index of the agent responsible for task $c_i$. Unlike random initialization or K-means, this
strategy directly exploits the geometric structure of the physical space to produce an initial partition with strong
spatial continuity, which not only shortens local route lengths but also stabilizes subsequent local search.

For local route optimization, the framework regards each cluster as an independent TSP with the corresponding depot as
both the start and end point, and employs a 2-opt algorithm based on the distance matrix to obtain a locally optimal
tour\cite{englert2014worst}. For the task set $\mathcal{C}_g$ associated with agent $g$, let $R_g$ denote the visiting order obtained by 2-opt,
and let \texttt{idxs} be the mapping from local indices to the global indices in the distance matrix. The route cost is
then given by
\begin{equation}
    L_g(\mathcal{C}_g)
    =
    \sum_{j=0}^{m_g}
    D_{\mathrm{idxs}[R_g[j]],\, \mathrm{idxs}[R_g[j+1]]}.
\end{equation}

Since each 2-opt move only involves a constant number of matrix lookups, the computational cost of evaluating
$L_g(\mathcal{C}_g)$ is effectively decoupled from the geometric computation, enabling near-linear scalability of local
routing.

To further improve global performance, the framework incorporates a neighborhood-based migration mechanism. For each
task point $c_i$, only the $M$ nearest depots are considered as potential migration targets, forming a candidate
neighborhood
\begin{equation}
    \mathcal{N}(c_i) = \mathrm{TopM}_g \,\|c_i - d_g\|_2.
\end{equation}
This step reduces the complexity of migration decisions from $O(nk)$ to $O(nM)$, substantially shrinking the search
space while preserving the effectiveness of local geometric structure.

When a task point $c_i$ is considered for migration, only the route costs of its current cluster and the candidate
target cluster need to be recomputed. Let the two clusters before migration be $\mathcal{C}_{\mathrm{cur}}$ and
$\mathcal{C}_{\mathrm{new}}$, respectively. After migrating $c_i$ from $\mathcal{C}_{\mathrm{cur}}$ to
$\mathcal{C}_{\mathrm{new}}$, the updated route costs are
\begin{equation}
    L'_{\mathrm{cur}} = L_{\mathrm{cur}}(\mathcal{C}_{\mathrm{cur}}\setminus\{i\}),
    \qquad
    L'_{\mathrm{new}} = L_{\mathrm{new}}(\mathcal{C}_{\mathrm{new}}\cup\{i\}).
\end{equation}
Combining these with the costs of all other unaffected clusters, we obtain the updated global cost indicators:
\begin{equation}
    \mathrm{new\_max} = \max_g L_g,
    \qquad
    \mathrm{new\_total} = \sum_g L_g.
\end{equation}
The acceptance of a migration is governed by a strict lexicographic criterion: the move is applied if and only if
\begin{equation}
    (\mathrm{new\_max},\, \mathrm{new\_total})
    <
    (\mathrm{max\_cost},\, \mathrm{total\_cost}),
\end{equation}
where $(\mathrm{max\_cost}, \mathrm{total\_cost})$ denotes the current global maximum route cost and total route cost,
respectively. This rule ensures that the algorithm primarily reduces the maximum route load and, subject to that,
further decreases the total cost, leading to a more balanced and efficient multi-agent task allocation. Owing to the
strict monotonicity of this lexicographic improvement, the local search process is guaranteed to converge to a stable
configuration within a finite number of iterations.

When no migration can further improve the lexicographic cost in a full outer iteration, the algorithm terminates.
A final 2-opt refinement is then applied to each cluster to obtain the final routes. Overall, the proposed framework
achieves a structural transformation from ``global ordering–evaluation'' to ``regional partitioning–local routing'',
effectively reducing the computational complexity of the multi-traveling salesman problem from exponential in the
permutation space to approximately linear in the partition space. As a result, scenarios involving thousands of tasks
and up to thousands of agents can be handled within a reasonable computation time.

The complete procedure of the proposed framework is summarized in Algorithm~\ref{alg:framework}.

\begin{algorithm}[t]
\caption{K-means Inspired Multi-Agent Task Allocation}
\label{alg:framework}
\begin{algorithmic}[1]

\STATE Build distance matrix: for all $p_i,p_j\in\mathcal{D}\cup\mathcal{C}$, set $D_{ij}=\|p_i-p_j\|_2$.

\STATE Initialize assignment: give each depot its nearest task, then assign remaining tasks by
$A_i=\arg\min_g\|c_i-d_g\|_2$.

\STATE Local routing: for each $g$, compute $L_g$ on $\mathcal{C}_g$ using 2-opt; record max\_cost and total\_cost.

\STATE Precompute neighborhoods: for each $c_i$, set $\mathcal{N}(c_i)=\mathrm{TopM}_g\|c_i-d_g\|_2$.

\REPEAT
  \FOR{each task $c_i$ in random order}
    \FOR{each $g\in\mathcal{N}(c_i)$}
      \STATE Evaluate moving $c_i$ by recomputing $L_g$ via 2-opt.
      \IF{$(\mathrm{new\_max},\mathrm{new\_total}) < (\mathrm{max\_cost},\mathrm{total\_cost})$}
        \STATE Accept migration and update costs.
      \ENDIF
    \ENDFOR
  \ENDFOR
\UNTIL no improvement

\STATE Final 2-opt on all clusters to obtain routes $R_g$.

\end{algorithmic}
\end{algorithm}

\section{Experimental Evaluation}

\subsection{Experimental Setup}

In order to comprehensively evaluate the proposed K-means inspired framework for solving large-scale MTSP, this section
provides a detailed description of the experimental environment, including the computing platform, dataset construction,
and baseline algorithm.

All experiments are conducted on a lightweight personal device equipped with an Apple M4 processor, using a
single-threaded Python implementation. The use of such a lightweight platform is intended to demonstrate that the
proposed framework can achieve high efficiency and good scalability under ordinary computational resources, without
relying on high-performance computing clusters or GPU acceleration.

For dataset construction, task points are distributed in a two-dimensional plane and all the points are uniformly
sampled within a given region to simulate widely dispersed task scenarios. To assess the scalability of
the algorithm, multiple problem sizes are considered. Specifically, the number of task
points $n$ is set to \(\{20, 50, 500, 5000\}\), and the number of agents $k$ is set to \(\{5, 10, 100, 1000\}\),
covering a wide range from small-scale to large-scale scenarios.

Furthermore, to thoroughly assess the performance of the proposed framework, a Genetic Algorithm (GA)\cite{lambora2019genetic} is selected as
the baseline method, and comparative experiments are carried out under identical hardware and task settings. As a
representative global-search heuristic, GA serves as a meaningful reference for evaluating the advantages of the
proposed method in terms of both computational efficiency and solution quality.

\subsection{Efficiency and Scalability Analysis}

\begin{figure*}[t]
\centering
\begin{subfigure}[b]{0.16\textwidth}
    \centering
    \includegraphics[width=\linewidth]{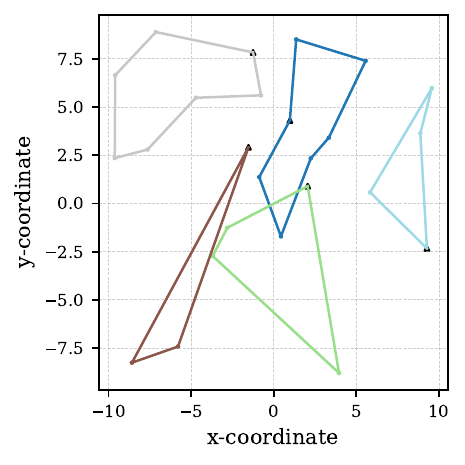}
    \caption{(a1) Routes}
    \label{fig:a1}
\end{subfigure}%
\begin{subfigure}[b]{0.16\textwidth}
    \centering
    \includegraphics[width=\linewidth]{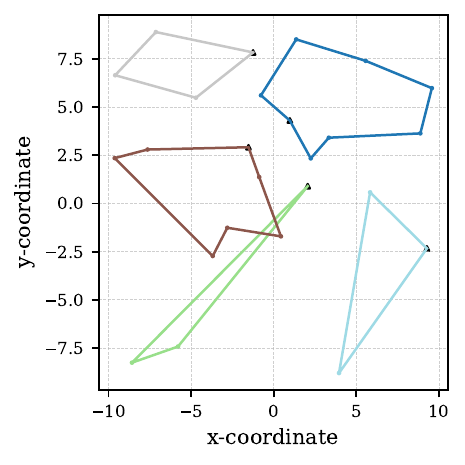}
    \caption{(a2) Routes}
    \label{fig:a2}
\end{subfigure}%
\begin{subfigure}[b]{0.16\textwidth}
    \centering
    \includegraphics[width=\linewidth]{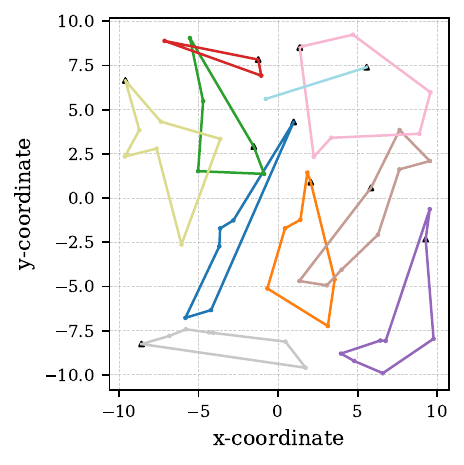}
    \caption{(b1) Routes}
    \label{fig:b1}
\end{subfigure}%
\begin{subfigure}[b]{0.16\textwidth}
    \centering
    \includegraphics[width=\linewidth]{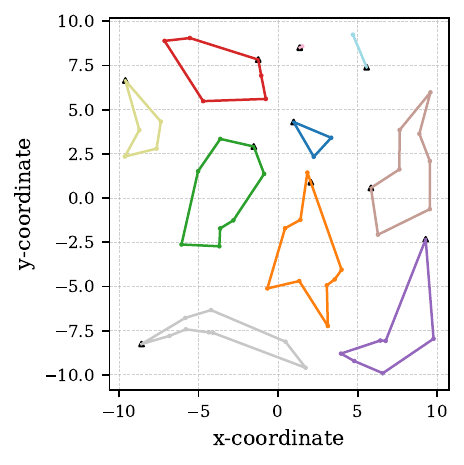}
    \caption{(b2) Routes}
    \label{fig:b2}
\end{subfigure}%
\begin{subfigure}[b]{0.16\textwidth}
    \centering
    \includegraphics[width=\linewidth]{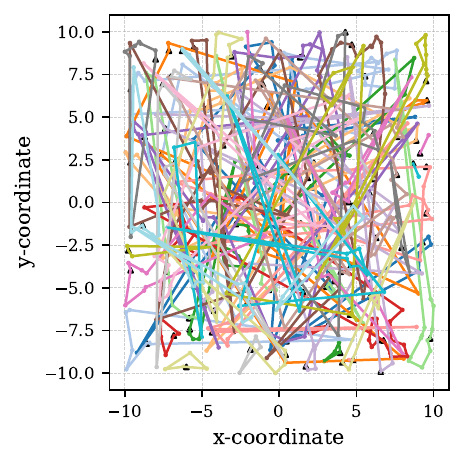}
    \caption{(c1) Routes}
    \label{fig:c1}
\end{subfigure}%
\begin{subfigure}[b]{0.16\textwidth}
    \centering
    \includegraphics[width=\linewidth]{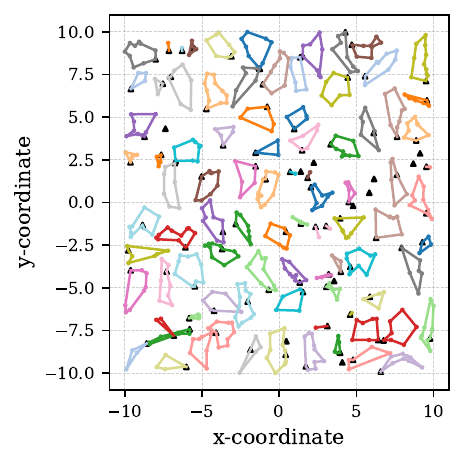}
    \caption{(c2) Routes}
    \label{fig:c2}
\end{subfigure}\\[0.6em] 

\begin{subfigure}[b]{0.16\textwidth}
    \centering
    \includegraphics[width=\linewidth]{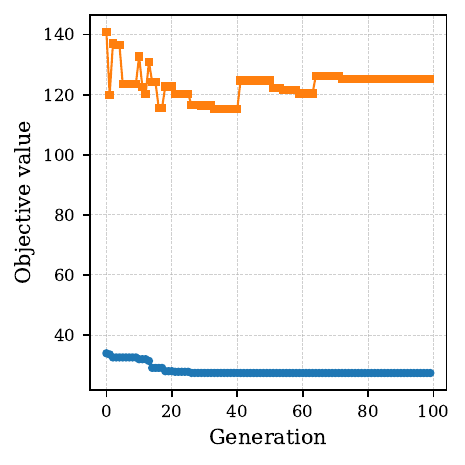}
    \caption{(d1) Convergence}
    \label{fig:d1}
\end{subfigure}%
\begin{subfigure}[b]{0.16\textwidth}
    \centering
    \includegraphics[width=\linewidth]{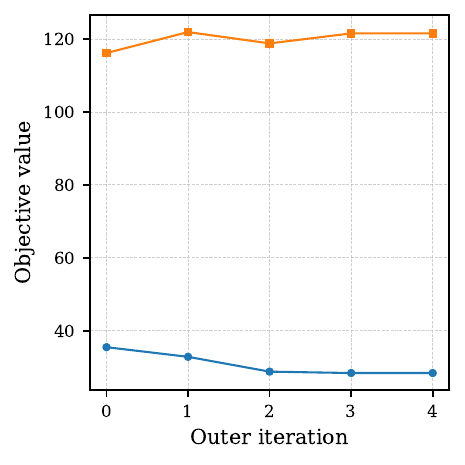}
    \caption{(d2) Convergence}
    \label{fig:d2}
\end{subfigure}%
\begin{subfigure}[b]{0.16\textwidth}
    \centering
    \includegraphics[width=\linewidth]{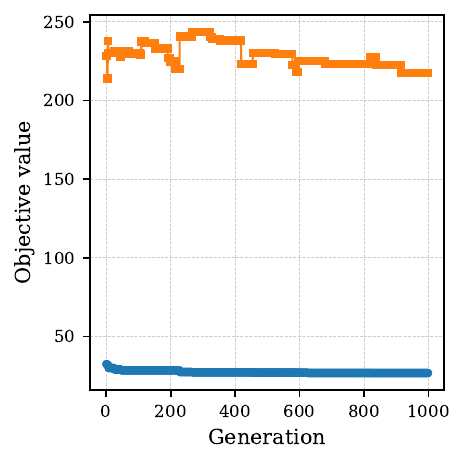}
    \caption{(e1) Convergence}
    \label{fig:e1}
\end{subfigure}%
\begin{subfigure}[b]{0.16\textwidth}
    \centering
    \includegraphics[width=\linewidth]{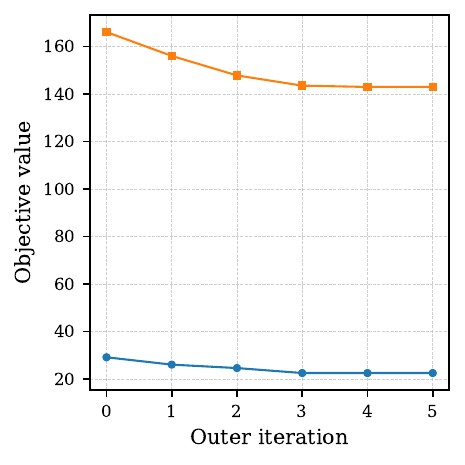}
    \caption{(e2) Convergence}
    \label{fig:e2}
\end{subfigure}%
\begin{subfigure}[b]{0.16\textwidth}
    \centering
    \includegraphics[width=\linewidth]{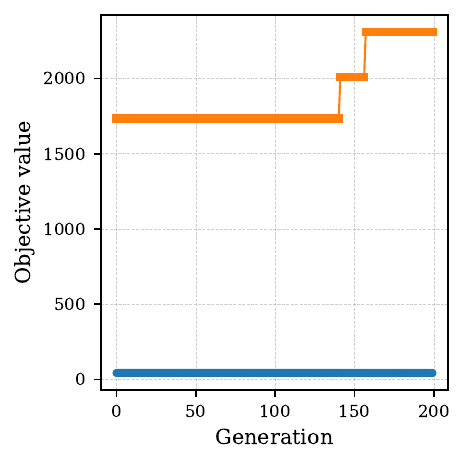}
    \caption{(f1) Convergence}
    \label{fig:f1}
\end{subfigure}%
\begin{subfigure}[b]{0.16\textwidth}
    \centering
    \includegraphics[width=\linewidth]{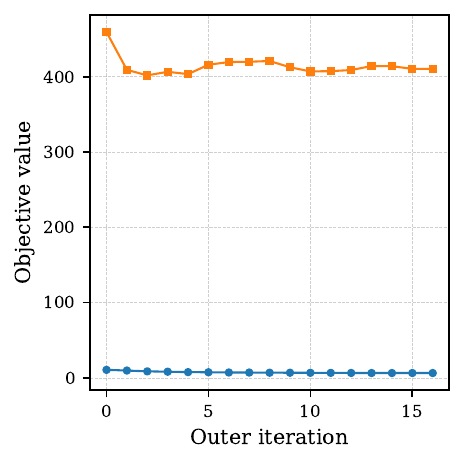}
    \caption{(f2) Convergence}
    \label{fig:f2}
\end{subfigure}

\caption{
Comparison between the proposed method and GA across different MTSP scales.
(a1)--(a2): 5 agents, 20 tasks;
(b1)--(b2): 10 agents, 50 tasks;
(c1)--(c2): 100 agents, 500 tasks.
(d1)--(d2), (e1)--(e2), (f1)--(f2) show the corresponding convergence curves.
}
\label{fig:algorithm_contrast}
\end{figure*}

To ensure a fair comparison, the Genetic Algorithm (GA) is implemented using the same distance matrix and 2-opt local
optimization as the proposed method, and both methods adopt an identical lexicographic objective: minimizing first the
maximum route length and then the total route length. Under the same random seed, three groups of depot and task point
coordinates are generated and shared by both algorithms. As illustrated in Fig.~\ref{fig:algorithm_contrast}~(a)--(c),
the two-dimensional allocation results of the two methods are visualized, where triangles represent depot (agent)
positions and circles represent task points. Figures~\ref{fig:algorithm_contrast}~(d)--(f) show the corresponding
convergence curves, and the quantitative comparison of the final solutions is summarized in Table~\ref{tab:comparison}.

In the scenario of Fig.~\ref{fig:algorithm_contrast}~(a1)--(a2), the number of agents is set to 5 and the number of
tasks to 20. GA achieves a maximum route length of 27.309 and a total route length of 125.238, with a population size
of 80, 100 iterations, and a mutation rate of 0.05. In contrast, the proposed method obtains a maximum route length of
28.379 and a total route length of 121.545, and converges in only 4 iterations. Since both methods adopt the same strict
lexicographic criterion, GA is slightly better in terms of solution optimality under this small-scale setting. This is
mainly because GA is able to explore a much larger solution space through global search, whereas the proposed method
relies on single-task migration moves and terminates early once no improving move can be found, making it more prone to
local optima in small instances. However, the number of iterations clearly indicates that the proposed method converges
extremely fast and can produce a feasible solution almost instantaneously in such small problem spaces.

In the scenario of Fig.~\ref{fig:algorithm_contrast}~(b1)--(b2), the number of agents is increased to 10 and the number
of tasks to 50. Due to the increased problem size, GA no longer converges quickly with the previous settings, and the
maximum number of iterations is therefore extended to 1000. GA converges at the 912-th iteration, yielding a maximum
route length of 26.770 and a total route length of 217.250, with a runtime of approximately 35.43 seconds. Under the
same instance, the proposed method attains a maximum route length of 22.523 and a total route length of 143.025, while
requiring only 5 iterations and 0.48 seconds of runtime, which is about 1.35\% of the GA runtime. In this regime, the
proposed approach achieves significantly better solution quality and much lower computational cost.

In the scenario of Fig.~\ref{fig:algorithm_contrast}~(c1)--(c2), the number of agents is further increased to 100 and
the number of tasks to 500. To maintain the feasibility of GA, the population size is reduced to 40; however, the
runtime per iteration remains extremely high, and GA fails to converge within a reasonable time limit. In contrast, the
proposed method converges after 16 iterations, with a runtime of 19.962 seconds, achieving a maximum route length of
6.536 and a total route length of 410.410. Under this medium-to-large scale setting, GA becomes practically unusable,
whereas the proposed framework still exhibits stable and efficient performance.

\begin{figure}[t]
    \centering
    \includegraphics[width=0.95\linewidth]{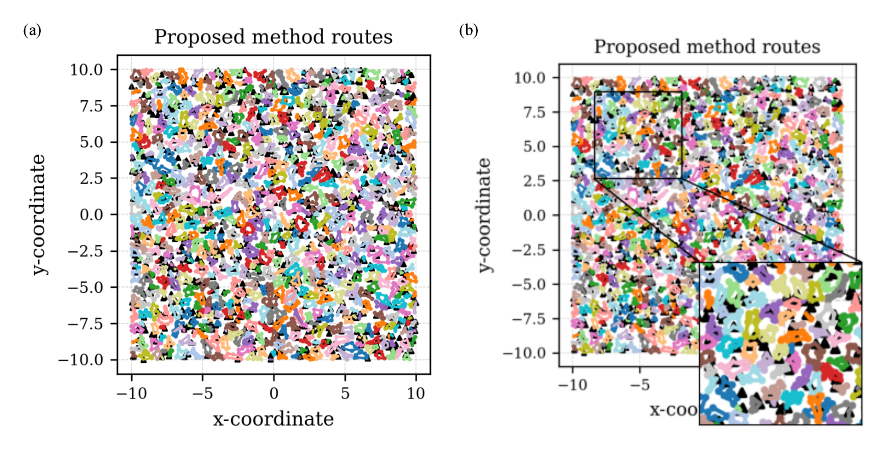}
    \caption{Performance of the proposed method in a large-scale scenario with 1000 agents and 5000 tasks.}
    \label{fig:PM1000k1_enlarge}
\end{figure}

Figure~\ref{fig:PM1000k1_enlarge} reports the results for an ultra-large-scale scenario with 1000 agents and 5000 tasks.
In this highly challenging setting with dense multi-agent coordination, GA is no longer able to run effectively, while
the proposed framework can still complete the task allocation within a limited time budget and converges after 44
iterations. This clearly demonstrates the strong scalability and robustness of the proposed K-means inspired framework
for large-scale MTSP instances.

\begin{table}[t]
\centering
\caption{Performance comparison between the proposed method and GA.}
\label{tab:comparison}
\begin{tabular}{lcccc}
\toprule
\textbf{Scenario} & \textbf{Method} & \textbf{Max Cost} & \textbf{Total Cost} & \textbf{Time (s)} \\
\midrule
5 agents, 20 tasks   & GA              & 27.309  & 125.238 & 1.182 \\
                     & Proposed        & 28.379  & 121.545 & 0.028 \\

10 agents, 50 tasks  & GA              & 26.770  & 217.250 & 35.432 \\
                     & Proposed        & 22.523  & 143.025 & 0.481 \\

100 agents, 500 tasks & GA             & --      & --      & Timeout \\
                      & Proposed       & 6.536   & 410.410 & 19.962 \\

1000 agents, 5000 tasks & GA           & --      & --      & Fail \\
                        & Proposed     & 2.767       & 1241.354      & 1237.215 \\
\bottomrule
\end{tabular}
\end{table}

\section{Conclusion}

This paper has proposed a K-means inspired framework for solving large-scale Multi-Traveling Salesman Problems (MTSP).
The framework restructures the traditional ``ordering--evaluation'' search paradigm into a two-stage pipeline of
``spatial clustering--local routing'', fundamentally reducing the search space and significantly lowering the number of
cost evaluations required during optimization. The method first constructs a spatially coherent initial partition using
a global distance matrix and geometric nearest-neighbor initialization, transforming task allocation into a
combinatorial optimization problem that can be efficiently solved within localized regions. Subsequently, by combining
fast 2-opt local TSP optimization with a candidate-neighborhood-based single-point migration strategy, the proposed
framework achieves rapid convergence with approximately linear computational complexity.

Experimental results demonstrate that the proposed approach exhibits notable advantages in terms of computational
efficiency and scalability. Under identical hardware settings, the proposed framework delivers orders-of-magnitude
speedup over a Genetic Algorithm (GA) baseline on medium-scale problems. For large-scale and ultra-large-scale
scenarios, the GA fails to converge within a reasonable time due to its substantial computational demands, whereas
the proposed method consistently yields stable and effective solutions without requiring parameter tuning. These
findings confirm the applicability and robustness of the proposed framework for large-scale multi-agent task allocation.

Nonetheless, certain limitations remain. Because the framework relies on a single-point migration mechanism, solution
diversity may be limited, and global optimality cannot be guaranteed in some scenarios. In ultra-large-scale cases,
the size of the candidate neighborhood strongly affects convergence speed, and the 2-opt local TSP solver may still
incur significant computational overhead when each agent is assigned a large number of tasks. Future work will explore
incorporating dynamic and kinematic constraints, extending the framework to heterogeneous agents, and leveraging
parallel computing and graph-based acceleration strategies to further enhance performance on even larger-scale MTSP
instances.

\bibliographystyle{unsrtnat}
\bibliography{references}

\end{document}